\documentclass[aps]{revtex4} 
\usepackage{natbib}
\usepackage{amsmath}
\usepackage{amssymb}
\begin{document}

\title{Solutions to the restricted three-body problem  with variable mass }
\author{Patricio S. Letelier}
\affiliation{Instituto de Matem\'atica, Estat\'istica e Computa\c c\~ao Cient\'ifica, Universidade Estadual de Campinas, 13083-859, Campinas, SP, Brazil}
\email{letelier@ime.unicamp.br}

\author{Tiago Amancio da Silva}
\affiliation{Instituto de F\'isica Gleb Wataghin, Universidade Estadual de Campinas, 13083-970, Campinas, SP, Brazil}
\email{amancio@ifi.unicamp.br}

\begin{abstract}
We look for  particular solutions to  the restricted three-body problem where the bodies are allowed to either lose or gain  mass to or from  a static atmosphere. In the case that all the masses are proportional to the same function of time,
we find analogous solution to the  five stationary solutions of  the usual restricted problem of constant masses:  the three collinear and the two triangular solutions, but  now the relative distance of the bodies changes with  time at the same rate. Under some restrictions, there are also coplanar, infinitely remote and ring solutions.
\end{abstract}
\keywords{three-body problem, variable masses}

\maketitle

\section{Introduction}

The study of the two-body problem of variable masses began practically with the work of \citet{gyl}, who wrote the differential equations of motion for the problem. The first integrable case to Gylden's equation was given by \citet{mes}, for a specific mass variation law. This mass variation law, and its following generalization \citep{mes2}, are known as \textit{Mestschersky laws}. After Mestschersky's contribution, the physical meaning of the problem became clear and it is known as \textit{Gylden-Mestschersky problem}.

 \citet{jeans}, by  studying the orbits of binary stars,  found a more general mass variation law that was  based on the relation between mass and luminosity of the stars presented by \citeauthor{eddington} in the same year. Mestschersky's laws are special cases of Jeans' law.

 \citet{gel} considered a different mass variation law. \cite{berkovic}  investigated the problem using a differential equation transformation method. Also a  number of approximate analytic solution were found, e.g., \cite{prieto}. \cite{luk3} studied the particular problem where the total mass is constant, which can be applied to conservative mass transfer in close binary systems \citep{luk4}.

The Gylden-Mestschersky problem can also be generalized  to include the restricted three-body problem.
In this approach, it is assumed that the two  heavier bodies  have their motion determined by the Gylden-Mestschersky equations. Thus, one have to deal only with the motion  of the third  body, which does not affect the main bodies motion. It was shown \citep{xx} that this problem presents particular solutions that are analogous to the stationary solutions of the classical problem of constant masses: the three collinear solutions $L_{1}$ to $L_{3}$ and the two triangular solutions $L_{4}$ and $L_{5}$. Many other particular solutions have also been found \citep{bekov,bekov2,luk,luk2}. Since then, further characteristics of this problem have been studied, for example, \citet{luk5,singh2}.

Besides the Gylden-Mestschersky problem, there are many different cases of two-body problems with variable mass \citep{razb}. These can be classified according to the presence or not of reactive forces, to the variation of the mass of just one or both of the bodies, to whether the bodies move in an inertial frame or not and so on.
 In this paper, we consider the specific case where the three bodies move in an inertial frame, within a static atmosphere, from which they absorb mass or to which they lose mass. We search for the five particular solutions analogous to the classic case ones, $L_1$ to $L_5$.

In the restricted three-body problem, the motion of the two primary body is determined \textit{a priori}. In the papers mentioned above, this motion is determined by the Gylden-Mestschersky problem \citep{mes2}, whereas in the present paper, the primary bodies move in a static atmosphere, from which they absorb or lose mass. Therefore, reactive forces are present. Fact not considered in the Gylden-Mestschersky problem.

\section{Particular solutions}

The equation of motion for a body whose mass depend on  time, $m\left(t\right)$, see for instance \citet{som}, is
\begin{eqnarray}
	\textbf{F}=m\dot{\textbf{v}}+\left(\textbf{v}-\textbf{u}\right)\dot{m},\label{pbm1}
\end{eqnarray}
where $\textbf{F}$ is the sum of all the forces acting on it and $\textbf{v}$ is its velocity, both measured in an inertial coordinate system. Also, $\textbf{u}$ is the velocity of the center of mass of the absorbed mass immediately before its union with the body (or of the ejected mass immediately after its ejection). The overdot denotes, as usual, derivation with respect to the time variable.

In the present context, there are two special cases of equation (\ref{pbm1}) to be considered. The first one is when the mass is ejected with the same velocity of the body at any moment ($\textbf{u}=\textbf{v}$), i.e.  mass ejection does not produce  reactive forces. This case can be used  to study the motion of a body ejecting mass isotropically (or radiating energy, since the total reactive momentum would be zero). In this case Eq. (\ref{pbm1}) takes the traditional form $\textbf{F}=m\dot{\textbf{v}}$. This case includes the Gylden-Mestschersky problem.

The second case, which we considered  in the present paper, is the one where $\textbf{u}=\textbf{0}$, i.e.  the particles are at rest in an inertial coordinate system. This case can be used to study the orbits of a star moving through a static atmosphere, whose particles attach or detach to the star as it moves. In this case, equation (\ref{pbm1}) reduces to
\begin{eqnarray}
	\textbf{F}=m\dot{\textbf{v}}+\dot{m}\textbf{v}=\frac{d}{dt}\left(m\textbf{v}\right).\label{pbm2}
\end{eqnarray}

According to (\ref{pbm2}), the equation of motion for a problem of two bodies of varying masses $m_{1}\left(t\right)$ and $m_{2}\left(t\right)$, exchanging mass with a static atmosphere surrounding them, is
\begin{eqnarray}
	&&\frac{d}{dt}\left(m_{1}\dot{\textbf{r}}_{1}\right)=-\frac{Gm_{1}m_{2}}{r^{3}}\textbf{r},\nonumber\\
	&&\frac{d}{dt}\left(m_{2}\dot{\textbf{r}}_{2}\right)=\frac{Gm_{1}m_{2}}{r^{3}}\textbf{r},\label{pbm3}
\end{eqnarray}
where $\textbf{r}_{1}$ and $\textbf{r}_{2}$ are the position vectors of the two bodies, $\textbf{r}=\textbf{r}_{1}-\textbf{r}_{2}$, $r=|\textbf{r}|$ and $G$ is the gravitational constant. We  note that the atmosphere around the bodies does not cause any drag forces on them. The only appreciable effect of this atmosphere is to work as a source or sink of mass for the bodies. From Eqs. (\ref{pbm3}) we get,
\begin{eqnarray}
	\frac{d}{dt}\left(m_{1}\dot{\textbf{r}}_{1}+m_{2}\dot{\textbf{r}}_{2}\right)=0.\label{pbm3.5}
\end{eqnarray}
The quantity in parenthesis can be set equal to zero in an  appropriated inertial frame. It follows, $\dot{\textbf{r}}_{1}=\left(m_{2}/M\right)\dot{\textbf{r}}$, where $M=m_{1}+m_{2}$, can be used to cast  equation (\ref{pbm3}) as
\begin{eqnarray}
	\frac{d}{dt}\left(\mu \dot{\textbf{r}}\right)=-\frac{G\mu M}{r^{3}}\textbf{r},\label{pbm4}
\end{eqnarray}
where $\mu\left(t\right)=m_{1}m_{2}/M$ is the reduced mass of the problem. In the classification table of the different two-body problems with variable masses by \citet{razb}, the present physical problem is listed as number 3. This problem also coincides mathematically with problems 14 and 15 (the later, known as Gelfgat-Omarov problem). Eq. (\ref{pbm4}) can be decomposed, in polar coordinates $\left(r,\theta\right)$, in the relations:
\begin{eqnarray}
	\frac{d}{dt}\left(\mu\dot{r}\right)&=&\frac{k^{2}}{\mu r^{3}}-\frac{GM\mu}{r^{2}},\label{pbm5}\\
	\mu r^{2}\dot{\theta}&=&k.\label{pbm6}
\end{eqnarray}
The coordinate $r$ is the distance between the primary bodies, $\theta$ is the angle between the line passing through their centers and a fixed line in the plane of motion, and $\dot{\theta}$ is the corresponding angular velocity. The constant $k$ is the total angular momentum of the system. Equation (\ref{pbm3.5}), together with the definition of the center of mass position vector, $M\textbf{R}_{cm}=m_{1}\textbf{r}_{1}+m_{2}\textbf{r}_{2}$, also yields the relation for the time dependence of $\textbf{R}_{cm}$
\begin{eqnarray}
	\dot{\textbf{R}}_{cm}=\frac{d}{dt}\left(\frac{m_{1}}{M}\right)\textbf{r}\label{pbm7}
\end{eqnarray}
which does not perform, in general, an inertial motion.

We shall  restrict ourselves to the case where the masses $m_{1}$ and $m_{2}$  vary arbitrarily, but their ratio remains constant. In this case  the center of mass of the system moves inertially. 
Following \citet{luk2}, the time dependence of the masses is described by a positive function $u\left(t\right)$:
\begin{eqnarray}
	m_{1}=m_{10}u\left(t\right),  m_{2}=m_{20}u\left(t\right),
	M=M_{0}u\left(t\right).
\end{eqnarray}
Furthermore, we have:
\begin{equation}
	\mu=\mu_{0}u\left(t\right).
\end{equation}
\label{pbm8}
The constants $m_{10}$, $m_{20}$, $M_{0}$ and $\mu_{0}$ are all positive, with $M_{0}=m_{10}+m_{20}$ and $\mu_{0}=m_{10}m_{20}/M_{0}$.

Now we introduce a barycentric rectangular coordinate system  that rotates comoving  with the main bodies, with angular velocity $\boldsymbol\omega=\omega\hat{z}$. The main bodies move in the $xy$ plane, in such a way the $x$ axis always passes through the center of the main bodies, at $x=x_{1}$ and $x=x_{2}$.

In such a coordinate system  the equations of motion for the third (lighter) body  are,
\begin{eqnarray}
	\frac{d}{dt}\left(m\dot{x}\right)&=&F_{x}+\dot{m}\omega y+2m\omega\dot{y}+m\dot{\omega}y+m\omega^{2}x,\nonumber\\
	\frac{d}{dt}\left(m\dot{y}\right)&=&F_{y}-\dot{m}\omega x-2m\omega\dot{x}-m\dot{\omega}x+m\omega^{2}y,\nonumber\\
	\frac{d}{dt}\left(m\dot{z}\right)&=&F_{z},\label{pbm9}
\end{eqnarray}
where $m\left(t\right)$ is the third body mass and $F_{x}$, $F_{y}$ and $F_{z}$ are the $x$, $y$ and $z$ components of the gravitational forces $\textbf{F}$, due to $m_{1}$ and $m_{2}$. It is instructive to
compare these two last equations with the corresponding equations for the motion of the third body in references like \citeauthor{bekov} (\citeyear{bekov} and \citeyear{bekov2}) and \citeauthor{luk} (\citeyear{luk} and \citeyear{luk2}).  Equations (\ref{pbm9}) have the additional terms $\dot{m}\omega y$ and $-\dot{m}\omega x$. These terms arise from the fact that the motion of the third body obeys equation (\ref{pbm4}). Equations (\ref{pbm9}) also appear in  \citet{bekov3}.

Following \citet{luk2}, we first write the distance between the primary bodies as $r=r_{0}R\left(t\right)$, where $R\left(t\right)$ is a positive function, so we have $\omega=\omega_{0}/uR^{2}$ (equations (\ref{pbm6}) and (\ref{pbm8})). Then, we define the mass parameter $\nu=m_{20}/M_{0}$ ($0<\nu\leq1/2$, without losing generality), and choose units such that $r_{0}=1$, $M_{0}=1$ and $\omega_{0}=1$. Therefore,
\begin{eqnarray}
	x_{1}&=&-\nu R\left(t\right),\nonumber\\
	x_{2}&=&\left(1-\nu\right)R\left(t\right).\nonumber
\end{eqnarray}

Furthermore, we shall assume  that the mass of the third body also varies as $m=m_{0}u\left(t\right)$, where $m_{0}$ is a positive constant. In the new units described above, the equations of motion (\ref{pbm9}) reduce to,
\begin{eqnarray}
	\frac{d}{dt}\left(u\dot{x}\right)&=&-\frac{Gu^{2}}{r_{1}^{3}}\left(1-\nu\right)\left(x+\nu R\right)\nonumber\\
	&&-\frac{Gu^{2}}{r_{2}^{3}}\nu\left(x-\left(1-\nu\right)R\right)+\nonumber\\
	&&+\frac{2\dot{y}}{R^{2}}-\frac{2y\dot{R}}{R^{3}}+\frac{x}{uR^{4}},\nonumber\\
	\frac{d}{dt}\left(u\dot{y}\right)&=&-\frac{Gu^{2}}{r_{1}^{3}}\left(1-\nu\right)y
	-\frac{Gu^{2}}{r_{2}^{3}}\nu y\nonumber\\
	&&-\frac{2\dot{x}}{R^{2}}+\frac{2y\dot{R}}{R^{3}}+\frac{y}{uR^{4}},\nonumber\\
	\frac{d}{dt}\left(u\dot{z}\right)&=&-\frac{Gu^{2}}{r_{1}^{3}}\left(1-\nu\right)z-\frac{Gu^{2}}{r_{2}^{3}}\nu z,\label{pbm10}
\end{eqnarray}
where $r_{1}=\sqrt{\left(x+\nu R\right)^{2}+y^{2}+z^{2}}$ and\\ $r_{2}=\sqrt{\left(x-\left(1-\nu\right) R\right)^{2}+y^{2}+z^{2}}$.  The constant $m_{0}$ cancels and it  does not appear in  equations (\ref{pbm10}).

We look for stationary solutions to  equations (\ref{pbm10}) of  the form  $x=\xi R\left(t\right)$, $y=\eta R\left(t\right)$ and $z=\zeta R\left(t\right)$, where $\xi$, $\eta$ and $\zeta$ are constants. By substituting these three relations for $x$, $y$ and $z$ into equations (\ref{pbm10}), we get
\begin{eqnarray}
	\xi\frac{d}{dt}\left(u\dot{R}\right)&=&-\frac{Gu^{2}}{R^{2}}\frac{\left(1-\nu\right)\left(\xi+\nu\right)}{\rho_{1}^{3}}\nonumber\\
	&&-\frac{Gu^{2}}{R^{2}}\frac{\nu\left(\xi+\nu-1\right)}{\rho_{2}^{3}}+\frac{\xi}{uR^{3}},\nonumber\\
	\eta\frac{d}{dt}\left(u\dot{R}\right)&=&-\frac{Gu^{2}}{R^{2}}\frac{\left(1-\nu\right)\eta}{\rho_{1}^{3}}\nonumber\\
	&&-\frac{Gu^{2}}{R^{2}}\frac{\nu\eta}{\rho_{2}^{3}}+\frac{\eta}{uR^{3}},\nonumber\\
	\zeta\frac{d}{dt}\left(u\dot{R}\right)&=&-\frac{Gu^{2}}{R^{2}}\frac{\left(1-\nu\right)\zeta}{\rho_{1}^{3}}-\frac{Gu^{2}}{R^{2}}\frac{\nu\zeta}{\rho_{2}^{3}},\label{pbm11}
\end{eqnarray}
where $\rho_{1}=\sqrt{\left(\xi+\nu\right)^{2}+\eta^{2}+\zeta^{2}}$ and\\ $\rho_{2}=\sqrt{\left(\xi+\nu-1\right)^{2}+\eta^{2}+\zeta^{2}}.$
The relative motion of $m_{1}$ and $m_{2}$ is given by $R\left(t\right)$. For this class of solutions  the geometrical configuration  of the system is  similar at any moment.
In order to find the time derivative of $u\dot{R}$, Eq. (\ref{pbm5}) is written  in the new units   as,
\begin{eqnarray}
	\frac{d}{dt}\left(u\dot{R}\right)=\frac{1}{uR^{3}}-\frac{Gu^{2}}{R^{2}}.\label{pbm12}
\end{eqnarray}

Finally, we obtain the form of equations (\ref{pbm11}) that allows us to find $\xi$, $\eta$ and $\zeta$
\begin{eqnarray}
	\xi-\frac{\left(1-\nu\right)\left(\xi+\nu\right)}{\rho_{1}^{3}}-\frac{\nu\left(\xi+\nu-1\right)}{\rho_{2}^{3}}&=&0, \nonumber\\
\eta\left(1-\frac{1-\nu}{\rho_{1}^{3}}-\frac{\nu}{\rho_{2}^{3}}\right)&=&0, \nonumber\\
\frac{\zeta}{uR^{3}}\left[1-GRu^{3}\left(1-\frac{1-\nu}{\rho_{1}^{3}}-\frac{\nu}{\rho_{2}^{3}}\right)\right]&=&0.\label{pbm13}
\end{eqnarray}
	To find the first two equations, we have divided by the nonzero  quantity $\frac{Gu^{2}}{R^{2}}$.
If we take $\zeta=0$ (solutions in the plane of motion of the primaries), the first two equations are exactly the corresponding equations for the restricted three-body problem of constant masses. This  means that all the stationary solutions of the constant masses problem are also present in the problem of variable masses discussed here, namely the three collinear solutions $L_{1}$ to $L_{3}$ (the three masses aligned) and the two triangular solutions $L_{4}$ and $L_{5}$ (the three masses at the vertices of an equilateral triangle), but now the relative distance of the bodies change with time at  the same rate.

In order to look for coplanar solutions, we take $\eta=0$. Then, equations (\ref{pbm13}) reduce to:
\begin{eqnarray}
	\xi-\frac{\left(1-\nu\right)\left(\xi+\nu\right)}{\rho_{1}^{3}}-\frac{\nu\left(\xi+\nu-1\right)}{\rho_{2}^{3}}&=&0, \nonumber\\
GRu^{3}\left(1-\frac{1-\nu}{\rho_{1}^{3}}-\frac{\nu}{\rho_{2}^{3}}\right)-1&=&0.\label{pbm14}
\end{eqnarray}
The existence of coplanar stationary solutions in the equations above requires
\begin{equation}
	GRu^{3}=constant=\kappa. \label{pbm15}
\end{equation}
In other words, the primary bodies must perform a particular motion, determined by a particular time variation rate for the masses masses. In \cite{luk2}, a similar restriction is found. Whereas in that work the restriction leads the masses to vary according to Mestschersky unified law, here, in general, one must solve numerically equations (\ref{pbm15}) and (\ref{pbm12}).  From equation (\ref{pbm14})  we have that  the possible values for $\kappa$ are restricted  by the relation 
$\kappa>1$.

Once equation (\ref{pbm15}) holds (with $\kappa>1$), equations (\ref{pbm14}) determine the existence of coplanar solutions \citep{luk}: $L_{6}$ and $L_{7}$, and also $L_{8}$ to $L_{11}$,
when  the value of $\kappa$ is less or equal to a certain value that depends on $\nu$. In the limiting case $\kappa\rightarrow 1$,  coplanar solutions do not exist, but the infinitely remote solutions $L_{\pm\infty}$ appear, with $\xi=\eta=0$ and $\zeta=\pm\infty$. Solutions $L_1$ to $L_5$ are still present.

\section{Ring solutions}

From our previous relations it is straightforward to verify the existence of ring ($L_{0}$) solutions. We consider now the collinear three-body problem with variable masses \citep{bekov2}. This is the particular case,  where the primary bodies move in a straight line passing through their centers. The motion of the primaries is still given by equation (\ref{pbm4}) and the ratio of their masses is still constant in time. We can obtain the radial equation of motion by setting $k=0$ in equations (\ref{pbm5}) and (\ref{pbm6}):
\begin{equation}
	\frac{d}{dt}\left(\mu\dot{r}\right)=-\frac{GM\mu}{r^{2}},\label{pbm16}
\end{equation}
with $\theta=constant$.

Since there is no need for the rotating frame of reference used before, we obtain the equations of motion for the third body by setting $\omega=0$ in equations (\ref{pbm9}), thus returning to the inertial frame. Hence, the analogous of equations (\ref{pbm10}) and (\ref{pbm11}) will present only the first two terms of the right-hand side of each of them. Furthermore, we notice, by symmetry, that it is enough to restrict ourselves to the plane $z=0$ and rotate the found solutions around the $x$-axis for all the spatial solutions. In the units used before, equation (\ref{pbm16}) is cast as:
\begin{eqnarray}
	\frac{d}{dt}\left(u\dot{R}\right)=-\frac{Gu^{2}}{R^{2}}\label{pbm17}
\end{eqnarray}
By substituting the above equation  in the analogous of equation (\ref{pbm11}), we get exactly the first two of equations (\ref{pbm13}). Therefore, we confirm the existence of solutions $L_{1}$ to $L_{3}$ and also solutions $L_{0}$ (a ring around the $x$-axis, generated by rotation of solutions $L_{4}$ and $L_{5}$).

\section{Conclusions}

The presence of a  static atmosphere (i.e., a source or sink of mass) in this three-body problem allows the center of mass and the relative motions of two main bodies to be different from that ones of the Gylden-Mestschersky problem. These motions can only be determined if the time variation of the masses is  known.

In summary,  we found  that our variable mass three body problem  
 has also  the  five Lagrange solutions of the classic three-body problem, $L_1$ to $L_5$, but now the relative distance of the bodies
change  with  time at the same rate. In the particular case in which the problem is collinear, triangular solutions $L_4$ and $L_5$ produce a ring solution $L_{0}$.

 The sufficient condition for the existence of these solutions is that  the ratio of their masses be constant  in time. Obviously, this restriction excludes many physically reasonable models for the mass variation, e.g., the ones where the rate of change of the masses depends on the masses themselves.
 
The coplanar solutions $L_6$ to $L_{11}$ and the infinitely remote solutions $L_{\pm\infty}$, that appear when the motion of the primaries is given by the Gylden-Mestschersky problem, are also present, under the very same additional conditions on the parameter $\kappa$ found in that case.

\acknowledgments
	T. S. Amancio thanks FAPESP and CAPES for financial support.
P.S. Letelier acknowledges FAPESP and CNPq for partial financial support.

\end{document}